\documentclass[a4paper,11pt]{article}
\usepackage{pos}

\title{The Shape of the Correlation Function}
\author*[a]{Jakub Cimerman}
\author[b,c]{Christopher Plumberg}
\author[a,d]{Boris Tom\'a\v{s}ik}

\affiliation[a]{Fakulta jadern\'a a fyzik\'aln\v{e} in\v{z}en\'yrsk\'a, \v{C}esk\'e vysok\'e u\v{c}en\'i technick\'e v Praze,\\
B\v{r}ehov\'a 7, 115~19 Praha 1, Czech Republic}

\affiliation[b]{Theoretical Particle Physics, Department of Astronomy and Theoretical Physics,\\ Lund University, S\"olvegatan 14A, SE-223 62 Lund, Sweden}

\affiliation[c]{Department of Physics, University of Illinois Urbana-Champaign,\\
1110 West Green Street, Urbana, IL 61801-3003, USA}

\affiliation[d]{Univerzita Mateja Bela,\\
Tajovsk\'eho 40, 974~01 Banská Bystrica, Slovakia}

\emailAdd{jakub.cimerman@fjfi.cvut.cz}
\abstract{The correlation function measured in ultrarelativistic nuclear collisions is strongly non-Gaussian. Using two different models we study which effects can influence its shape and how much. In particular,
we focus on the parametrizations expressed with the help of Lévy-stable distributions. We show that
the Lévy index may deviate substantially from 2 due to non-critical effects such as resonance decays, event-by-event fluctuations and functional dependence on $Q_{\LI}$ or similar. We also study the corrections including the first-order Lévy expansion.
}

\FullConference{%
  40th International Conference on High Energy physics - ICHEP2020\\
  July 28 - August 6, 2020\\
  Prague, Czech Republic (virtual meeting)
}

\def\LI{\mathrm{LI}}
\def\LBI{\mathrm{LBI}}
\def\ev{\mathrm{ev}}

\begin{document}
\maketitle

\section{Introduction}

Correlation femtoscopy \cite{Lisa:2005dd} provides us an information about the sizes of the homogeneity region of the fireball in relativistic heavy-ion collisions. The two-particle correlation functions were fitted with a Gaussian form at first, but the real shape turned out to be strongly non-Gaussian and thus it is better described by a Lévy-stable distribution. Recent experimental results \cite{Adare:2017vig} reports a value of Lévy index well below 2. There are arguments that the value of Lévy index equal to 0.5 may identify matter produced at the critical endpoint of the QCD phase diagram. Nevertheless, there are some non-critical effects which can affect the value of the Lévy index significantly

%%%%%%%%%%%%%%%%%%%%%%%%%%%%%%%%%%%%%%%%%%%%%%%%%%%%%%%%%%%%%%%%%%%%%%%%%%%%%%%%%%%%%%%%%%%%%%%

\section{HBT Formalism}

Two-particle correlation function probes the momentum-space structure of correlations between pairs of particles produced in heavy-ion collisions. In this work, we focus on correlations between charged pion pairs. We use the correlation function in the form
\begin{equation}
C(q,K)-1\approx\dfrac{\Big|\int \mathrm{d}^4xS(x,K) e^{iqx} \Big|^2}{\left(\int \mathrm{d}^4xS(x,K)\right)^2},
\end{equation}
where $q=p_1-p_2$ is the momentum difference, $K=\frac{1}{2}(p_1+p_2)$ is the average momentum and $S(x,K)$ is the emission function which describes the probability that a particle with momentum $p$ is emitted from position $x$.
Also the smoothness approximation $K\approx p_1 \approx p_2$ is assumed.

%Since the Gaussian parametrization often does not adequately describe the experimentally measured correlation function, we focused on L\'evy parametrization of the correlation function 
The L\'evy parametrization is more general
\begin{equation}
C_L(\vec{q},\vec{K}) = 1+ \lambda'(\vec{K}) \exp \left[ - \Bigg| \sum_{i,j=o,s,l} R_{ij}^{'2}(\vec{K}) q_i q_j \Bigg| ^{\alpha/2} \right].
\label{eq-3dlevy}
\end{equation}
The parameters $\lambda'$ and $R_{ij}^{'2}$ are analogous to those used in the Gaussian parametrization, but have no direct correspondence with the source widths. The L\'evy index $\alpha$  controls the form of the distribution used to approximate correlation function: for  $\alpha=2$ L\'evy distribution becomes  Gaussian, while for  $\alpha=1$ it becomes exponential. The one-dimensional L\'evy parametrization has form
\begin{equation}
C_L(Q) = 1 + \lambda' \exp (-|R'Q|^\alpha).
\label{eq-1dlevy}
\end{equation}

%%%%%%%%%%%%%%%%%%%%%%%%%%%%%%%%%%%%%%%%%%%%%%%%%%%%%%%%%%%%%%%%%%%%%%%%%%%%%%%%%%%%%%%%%%%%%%%%%%%%%%%%%%%%%%%%%%%%%%%%

\section{Effects Leading to Non-Gaussian Behaviour}

We focused on three effects which can lead to non-Gaussian behaviour. The first is event averaging. Each event differs from the rest due to random fluctuations of size, geometric and dynamical anisotropies, and so on. The correlation function is  conventionally  averaged over a large number of different events. Its formula  thus must be replaced by \cite{Plumberg:2013nga,Plumberg:2015mxa}
\begin{equation}
C(q,K)\approx 1+\dfrac{\left\langle \Big|\int \mathrm{d}^4xS(x,K)e^{iqx}\Big|^2\right\rangle_{\ev}}{\left\langle\left(\int \mathrm{d}^4xS(x,K)\right)^2\right\rangle_{\ev}}.
\end{equation}

The second effect is due to the use of a one-dimensional projection of the relative momentum. The correlation function is then a function of a single scalar quantity: either the Lorentz-invariant variable
\begin{equation}
%Q_{LI}^2 \equiv 
Q^2_{\LI}= -q^\mu q_\mu = \vec{q}\cdot \vec{q}-(q^0)^2
\label{eq-qli}
\end{equation}
or a longitudinally boost-invariant one \cite{Adare:2017vig}
\begin{equation}
Q_{\LBI}^2=\sqrt{(p_{1x}-p_{2x})^2+(p_{1y}-p_{2y})^2+\frac{(p_{1z}E_2-p_{2z}E_1)^2}{K_0^2-K_l^2}}.
\label{eq-qlbi}
\end{equation}

The third effect we studied is the impact of resonance decays on the L\'evy index. Different resonances contribute to the correlation function with different lengthscales and timescales, while the Gaussian function is given by only a single lengthscale. Therefore, the correlation function must deviate from a Gaussian form once resonance effects are included.

%%%%%%%%%%%%%%%%%%%%%%%%%%%%%%%%%%%%%%%%%%%%%%%%%%%%%%%%%%%%%%%%%%%%%%%%%%%%

\section{Models}

To show that our results are not artifacts due to specific model, we  use two different models. Firstly, the blast-wave model \cite{Retiere:2003kf}, which describes an expanding locally thermalised fireball. It is characterized by the emission function
\begin{equation}
S(x,p)\mathrm{d}^4x  =\frac{m_t\cosh (\eta- Y)}{(2\pi)^3}\mathrm{d}\eta \mathrm{d}x\mathrm{d}y\frac{\tau \mathrm{d}\tau}{\sqrt{2\pi}\Delta\tau} \exp \left(-\frac{(\tau-\tau_0)^2}{2\Delta\tau^2}\right) \exp\left(-\frac{E^\ast}{T}\right)\Theta \left(1-\overline{r}\right),
\end{equation}
where $\Theta \left(1-\overline{r}\right)$ is Heaviside step function, $E^\ast = p_\mu p^\mu$ is the energy in the rest frame of the fluid and $\overline{r}=\frac{r}{R(\theta)}$ is a scaled radius of fireball in transverse plane. 

We used DRAGON \cite{Tomasik:2008fq} to generate events, which is a Monte Carlo event generator based on the blast-wave model with added resonance decays. For this study we generated sets of 50,000 events with parameters set to: temperature $T=120\:\mathrm{MeV}$, the average transverse radius $R_0=7\:\mathrm{fm}$, freeze-out time $\tau_{\mathrm{fo}}=10\:\mathrm{fm/}c$, the strength of the transverse expansion $\rho_0 = 0.8$. CRAB \cite{crab} was then used to generate correlation functions from these events.

The second is the  hydrodynamic model iEBE-VISHNU \cite{Song:2007ux, Shen:2014vra}. It is a 2+1-D hydrodynamic simulation with  M\"uller-Israel-Stewart  equations and Glauber MC initial conditions. We generated 1,000 events of $0-10\%$ Au+Au collisions at 200$A$ GeV with  freeze-out temperature $120\:\mathrm{MeV}$ and $\eta/s=0.08$. To compute the HBT correlation function we used the HoTCoffeeh code \cite{Plumberg:2016sig}, which directly evaluates Cooper-Frye integrals over the freeze-out surface on an event-by-event basis. 

%%%%%%%%%%%%%%%%%%%%%%%%%%%%%%%%%%%%%%%%%%%%%%%%%%%%%%%%%%%%%%%%%%%%%%%%%%%%

\section{Results}

\begin{figure}[t!]
\centering
\includegraphics[width=0.4\textwidth]{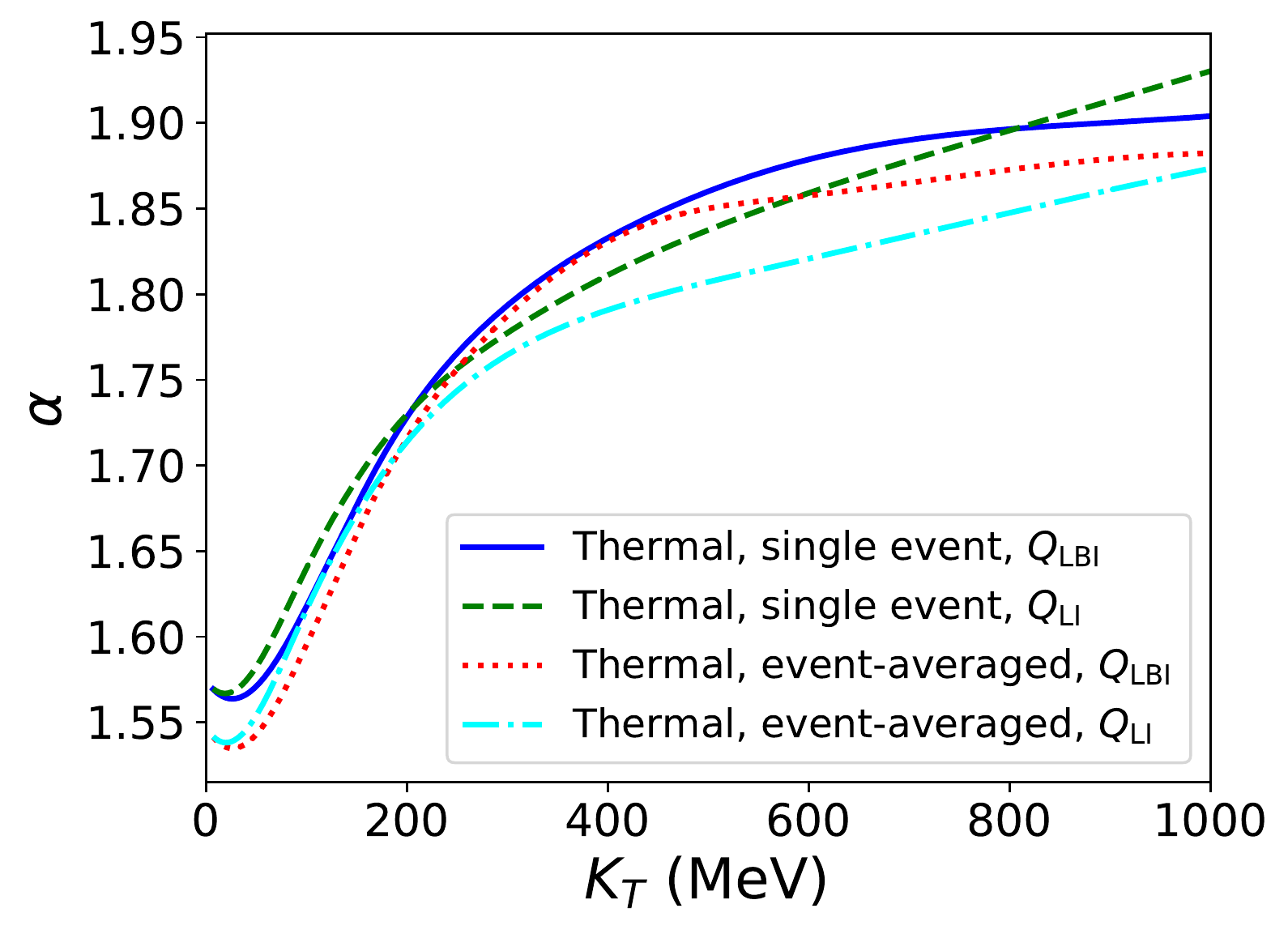}
\includegraphics[width=0.4\textwidth]{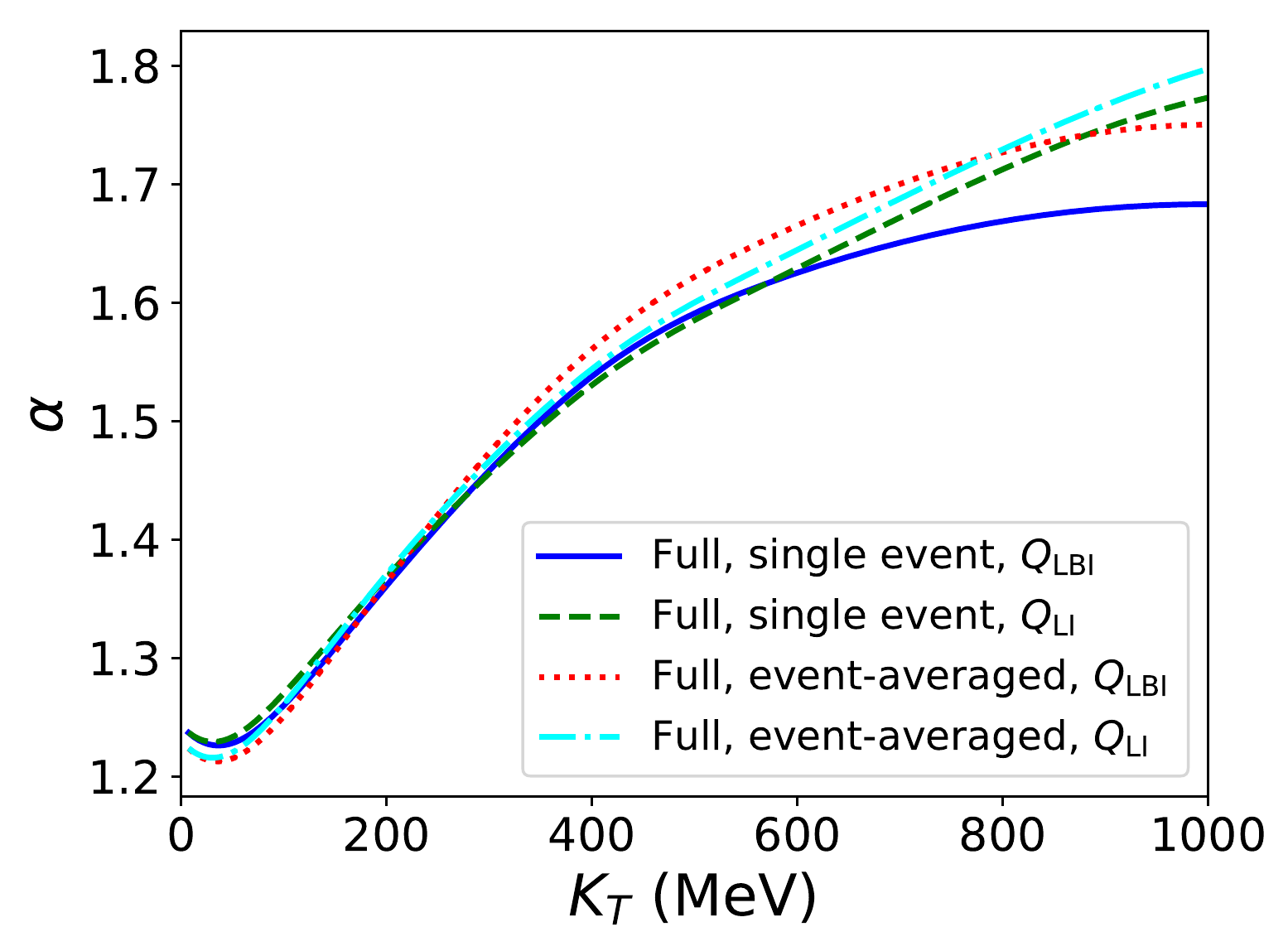}
\caption{A comparison of $\alpha (K_T)$ with and without different non-Gaussian effects in the hydrodynamic model: with and without event averaging and for different choices of $Q$ (solid blue and dashed green vs. dotted red and dash-dotted cyan). The comparison is made both for thermal pions only (left panel) and for the full thermal and resonance contributions added together (right panel).}
\label{fig1}
\end{figure}

We calculated 1D and 3D correlation functions and obtained  the L\'evy index with a 1D fit using Eq. \eqref{eq-1dlevy} or a 3D fit using Eq. \eqref{eq-3dlevy}. We focus on the $K_T$-dependence of the L\'evy index. First, we used the hydrodynamic model to check the relative importance of several of the effects discussed above. In Figure \ref{fig1} we see the impact of three of them: correlation function with resonances vs. without resonances, single event  vs. event-averaged, and finally $Q_{\LI}$ vs. $Q_{\LBI}$.

%%%%%%%%%%%%%%%%%%%%%%%
%%%%%%%%%%%%%%%%%%%%%%%

From this figure, we can say that the latter two effects do not affect the L\'evy index significantly. We find that the inclusion of resonances reduces the value of the L\'evy index by 0.2-0.3. Nevertheless, as we show below, the largest effects are concentrated mainly at low $K_T$ and are due to the use of the 1D projection of the relative momentum.

\begin{figure}[b!]
\centering
\includegraphics[width=0.4\textwidth]{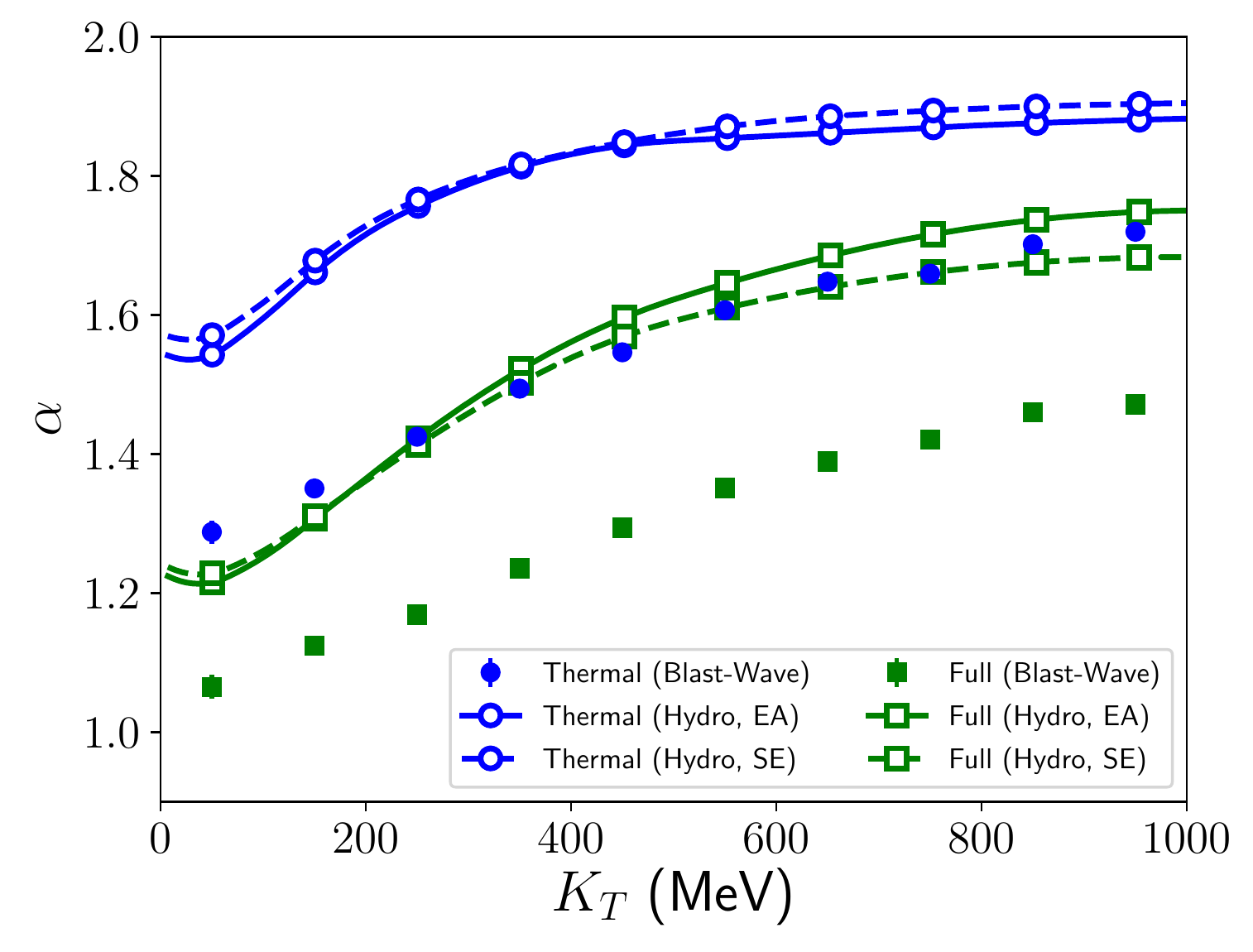}
\includegraphics[width=0.4\textwidth]{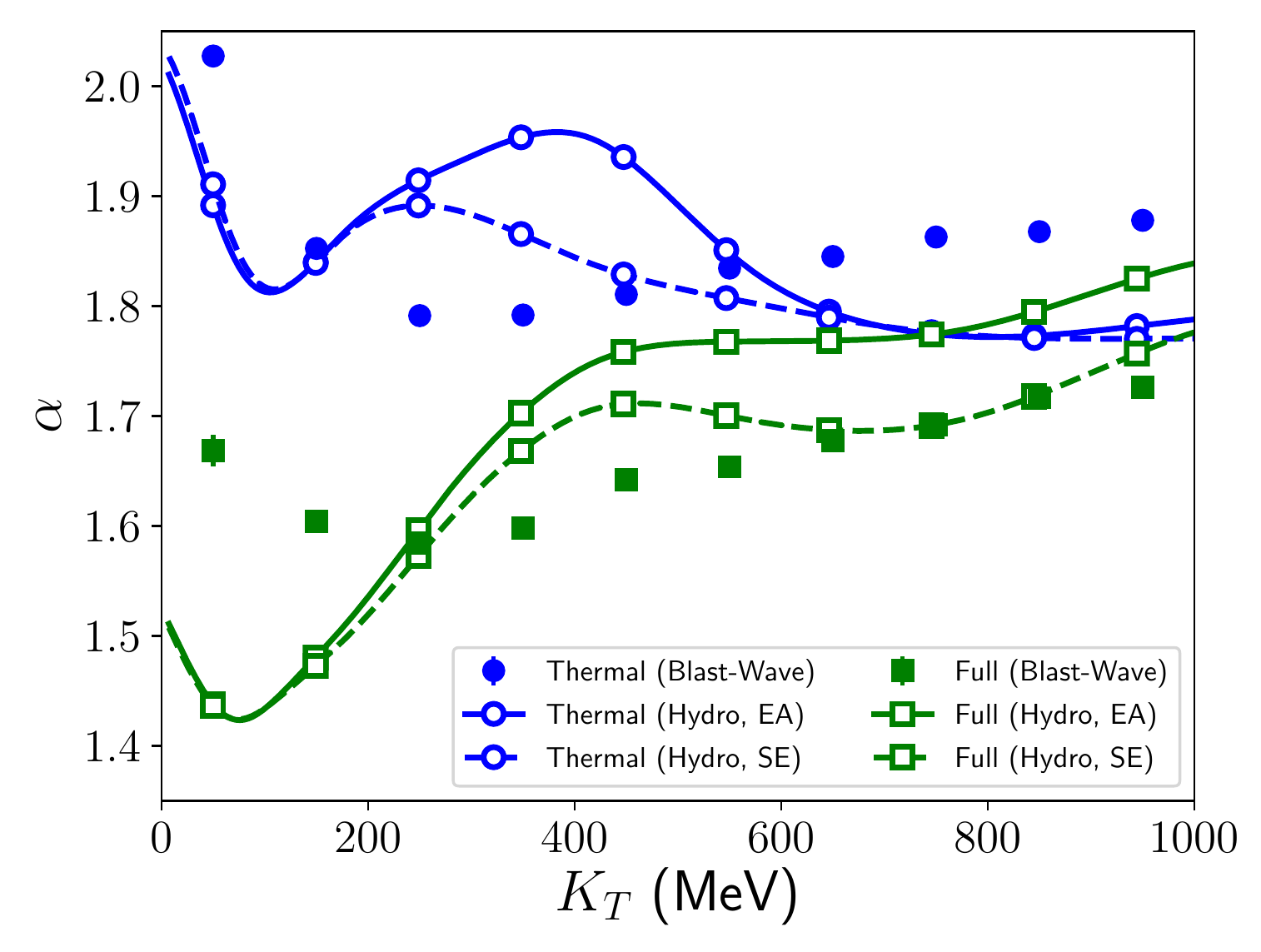}
\caption{The L\'evy index of the 1D fit to the correlation function in $Q_{\LI}$ (left) and the L\'evy index of the 3D fit to the correlation function according to Eq. \eqref{eq-3dlevy} (right). The blue circles show results from a source without resonances, while the green squares show results from a source with resonances. The solid points correspond to the blast-wave model and the open points represent the hydrodynamic results for event-averaged (solid) and single-event (dashed) correlation functions.}
\label{fig3}
\end{figure}

To estimate the model-independent impact of resonances on the L\'evy index, we calculated its $K_T$-dependences using both our models (Figure \ref{fig3}). This plot shows that, regardless of the model used in calculations, resonances reduce the value of the L\'evy index by $\sim$0.2. The figure contains also the comparison of the 1D (left) and 3D (right) Lévy fit of the correlation function. By comparing them it can be seen that the 1D projection cause a large decrement of Lévy index. 

%%%%%%%%%%
%%%%%%%%%%%%

\begin{figure}[t]
\centering
\includegraphics[width=0.4\textwidth]{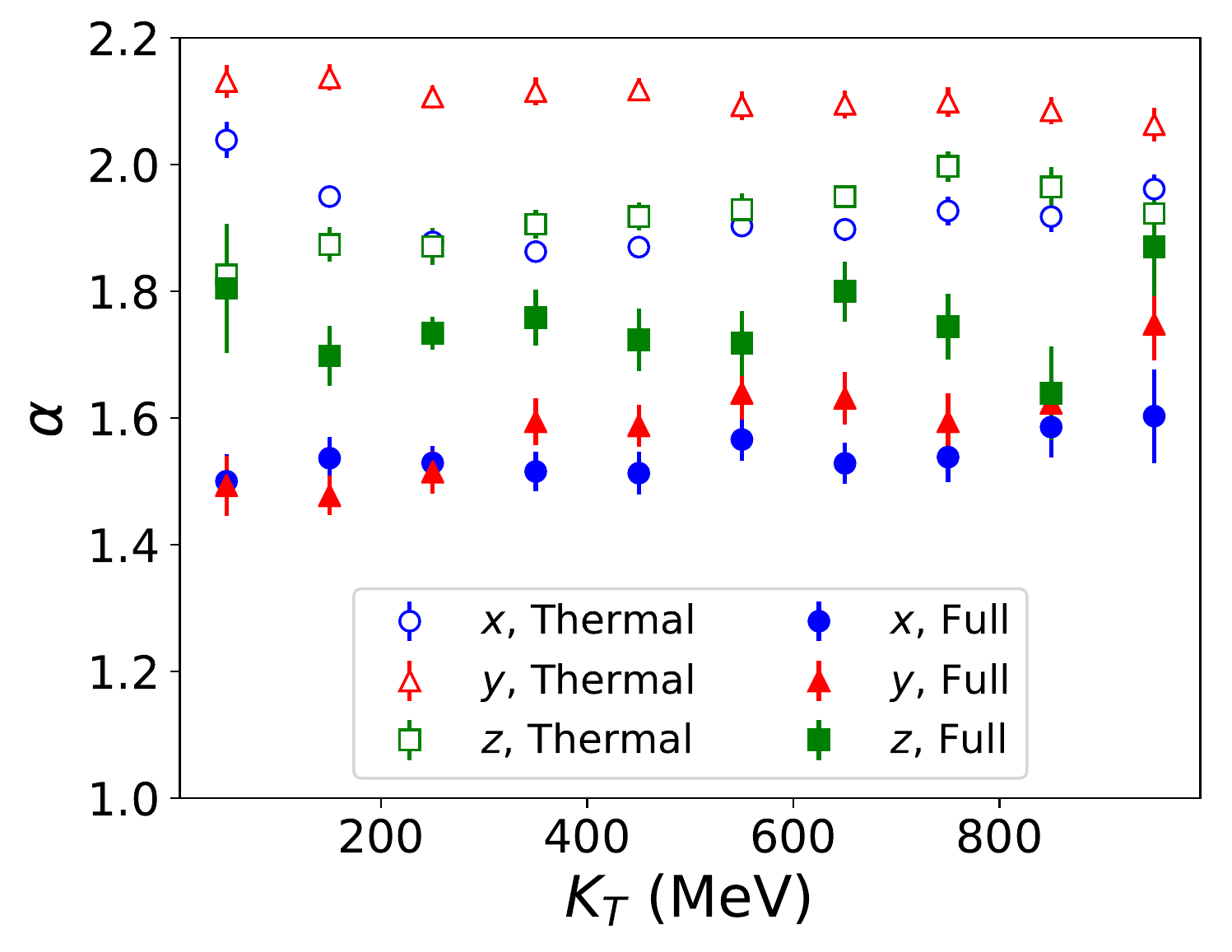}
\includegraphics[width=0.4\textwidth]{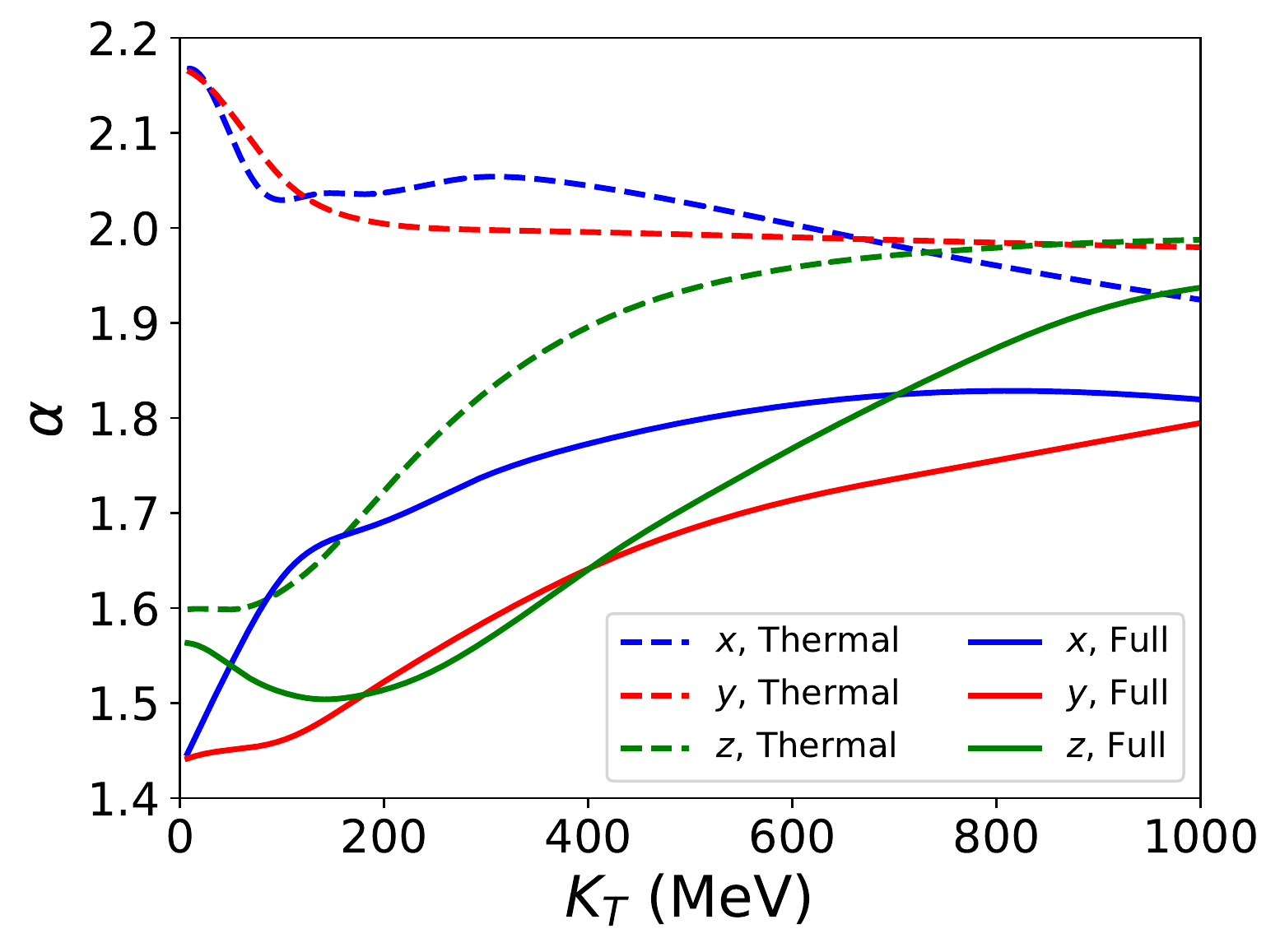}
\caption{The L\'evy index of the 1D fits to the correlation function in $\vec{q}$ along different axes, with or without resonances. In this figure $x$ corresponds to outward, $y$ to sideward and $z$ to longitudinal direction. Left panel: blast-wave model. Right panel: hydrodynamics.}
\label{fig4}
\end{figure}

To find out why does the 1D projection affect L\'evy index so significantly we  look at different directions of 3D correlation function (Figure \ref{fig4}). These plots show, that while the correlation function behaves similarly in outward and sideward direction, the $K_T$-dependence in longitudinal direction behaves differently. Moreover, it seems that the resonances do not affect the correlation function in the longitudinal direction as much as in the transverse plane. 

%%%%%%%%%%%%%%%%%%%%%%%%%%%%%%%%%%%%%%%%%%%%%%%%%%%%%%%%%%%%%%%%%%%%%%%%%%%%%%%%%%%%%%%%%%

\section{Corrections}

\begin{figure}[h]
    \centering
    \includegraphics[width=0.47\textwidth]{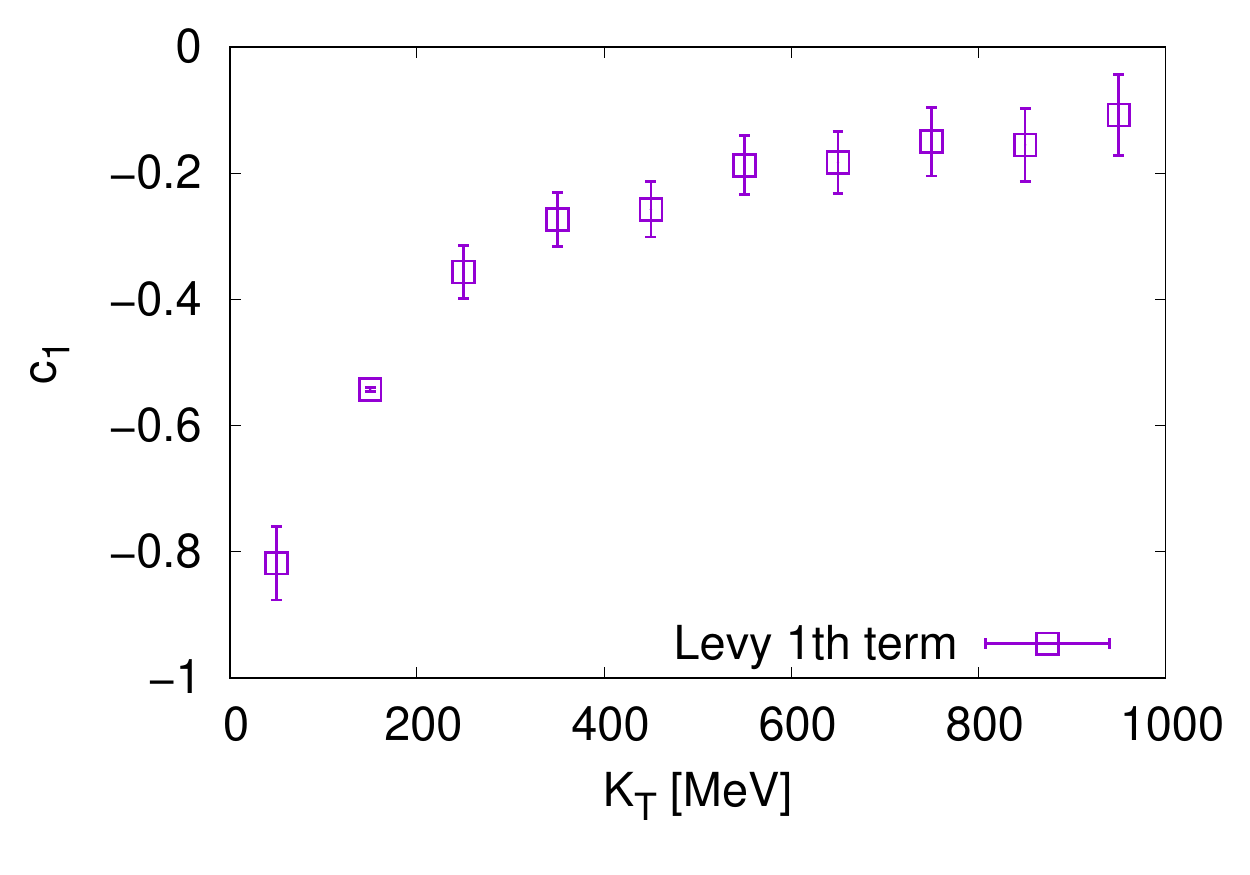}
    \caption{The first-order correction coefficient $c_1$ as a function of $K_T$ obtained from 3D fit to the correlation function according to Eq. \eqref{eq-3dlevy}.}
    \label{fig:c1}
\end{figure}

Even Lévy parametrization cannot describe our correlation functions perfectly. Thus we need to study the corrections to higher orders. We decompose the data into Lévy expansion up to first order \cite{Csorgo:2000pf}
\begin{equation}
    C(Q) = 1+ \lambda e^{-R^\alpha Q^\alpha} \left[1+c_1L_1(Q|\alpha)\right],
\end{equation}
where $c_1$ is a complex expansion coefficient and $L_1$ is the Lévy polynomial given by the formula
\begin{equation}
    L_1(t|\alpha)=\frac{1}{\alpha}\left[ \Gamma\left(\frac{1}{\alpha}\right)t-\Gamma\left(\frac{2}{\alpha}\right) \right].
\end{equation}
Unfortunately, we found such fits very unstable. We have been able to obtain the coefficient $c_1$ but not the higher order corrections. The dependence of $c_1$ on $K_T$ is shown in Figure \ref{fig:c1} for a 3D fit to a 3D correlation function. It can be seen that the $c_1$ is closer to 0 for higher $K_T$, but the general conclusion is that the first-order is not negligible and thus our correlation functions are neither Lévy-shape.

%%%%%%%%%%%%%%%%%%%%%%%%%%%%%%%%%%%%%%%%%%%%%%%%%%%%%%%%%%%%%%%%%%%%%%%%%%%%%%%%%%%%%%%%%%%

\section{Conclusions}

We have shown that the shape of the correlation function, as well as the index of the Lévy-stable parametrization fitted to the correlation function, may be influenced by a variety of different mechanisms. All our results show that the L\'evy index may deviate substantially from the value of 2 due to non-critical effects. Not all of the studied effects have  significant influence, but two of them are found to cause notable deviations. The first arise from the projection of the 3D relative momentum $\vec{q}$ onto a scalar $Q$, and the second arise from the inclusion of resonance decays. Since we used two different models, these results appear to be robust and not merely artifacts of the models we have used. For this reason, the conclusions presented here may be regarded as model-independent.

\subsubsection*{Acknowledgements}

BT acknowledges support from VEGA 1/0348/18 (Slovakia). CP gratefully acknowledges funding from the CLASH project (KAW 2017-0036) and the US-DOE Nuclear Science Grant No. DE-SC0019175, as well as the use of computing resources from both the Minnesota Supercomputing Institute (MSI) at the University of Minnesota and the Ohio Supercomputer Center \cite{OSC} which contributed to the research results reported within this proceedings.

\end{document}